# Direct measurement of the upper critical field in a cuprate superconductor

G. Grissonnanche[1], O. Cyr-Choinière[1], F. Laliberté[1], S. René de Cotret[1],

A. Juneau-Fecteau[1], S. Dufour-Beauséjour[1], M.-È. Delage[1], D. LeBoeuf[1,†], J. Chang[1,‡],

B. J. Ramshaw[2], D. A. Bonn[2,3], W. N. Hardy[2,3], R. Liang[2,3], S. Adachi[4], N. E. Hussey[5],

B. Vignolle[6], C. Proust[3,6], M. Sutherland[7], S. Krämer[8], J.-H. Park[9], D. Graf[9],

N. Doiron-Leyraud[1] & Louis Taillefer[1,3]

*1 Département de physique & RQMP, Université de Sherbrooke, Sherbrooke, Québec J1K 2R1, Canada*

*2 Department of Physics & Astronomy, University of British Columbia, Vancouver, British Columbia V6T 1Z1, Canada*

*3 Canadian Institute for Advanced Research, Toronto, Ontario M5G 1Z8, Canada*

*4 Superconductivity Research Laboratory, ISTC Tokyo 135-0062, Japan*

*5 H. H. Wills Physics Laboratory, University of Bristol, Bristol BS8 1TL, UK*

*6 Laboratoire National des Champs Magnétiques Intenses, Toulouse 31400, France*

*7 Cavendish Laboratory, University of Cambridge, Cambridge CB3 0HE, UK*

*8 Laboratoire National des Champs Magnétiques Intenses, Grenoble, France*

*9 National High Magnetic Field Laboratory, Tallahassee, Florida, USA*

† Present address : Laboratoire National des Champs Magnétiques Intenses, Grenoble, France.

‡ Present address : École Polytechnique Fédérale de Lausanne, CH-1015 Lausanne, Switzerland.

**The upper critical field $H_{c2}$ is a fundamental measure of the pairing strength, yet there is no agreement on its magnitude and doping dependence in cuprate superconductors[1,2,3]. We have used thermal conductivity as a direct probe of $H_{c2}$ in the cuprates $YBa_2Cu_3O_y$ and $YBa_2Cu_4O_8$ to show that there is no vortex liquid at $T = 0$, allowing us to use high-field resistivity measurements to map out the doping dependence of $H_{c2}$ across the phase diagram. $H_{c2}(p)$ exhibits two peaks, each located at a critical point where the Fermi surface undergoes a transformation[4,5]. The condensation energy obtained directly from $H_{c2}$, and previous $H_{c1}$ data[6,7], undergoes a 20-fold collapse below the higher critical point. These data provide quantitative information on the impact of competing phases in suppressing superconductivity in cuprates.**

In a type-II superconductor at $T = 0$, the onset of the superconducting state as a function of decreasing magnetic field $H$ occurs at the upper critical field $H_{c2}$, dictated by the pairing gap $\Delta$ through the coherence length $\xi_0 \sim v_F / \Delta$, via $H_{c2} = \Phi_0 / 2\pi\xi_0^2$, where $v_F$ is the Fermi velocity and $\Phi_0$ is the flux quantum. $H_{c2}$ is the field below which vortices appear in the sample. Typically, the vortices immediately form a lattice (or solid) and thus cause the electrical resistance to go to zero. So the vortex-solid melting field, $H_{vs}$, is equal to $H_{c2}$. In cuprate superconductors, the strong 2D character and low superfluid density cause a vortex liquid phase to intervene between the vortex-solid phase below $H_{vs}(T)$ and the normal state above $H_{c2}(T)$ (ref. 8). It has been argued that in underdoped cuprates there is a wide vortex-liquid phase even at $T = 0$ (refs. 2,9,10,11), so that $H_{c2}(0) >> H_{vs}(0)$, implying that $\Delta$ is very large. So far, however, no measurement on a cuprate superconductor has revealed a clear transition at $H_{c2}$, so there are only indirect estimates (refs. 1,2,3) and these vary widely (see Fig. S1 and associated discussion). For example, superconducting signals in the Nernst effect[2] and the magnetization[11] have been tracked to high fields, but it is difficult to know whether these are due to vortex-like excitations below $H_{c2}$ or to fluctuations above $H_{c2}$ (ref. 3).

To resolve this question, we use the fact that electrons are scattered by vortices, and monitor their mobility as they enter the superconducting state by measuring the thermal conductivity $\kappa$ of a sample as a function of magnetic field $H$. In Fig. 1, we report our data on two cuprate superconductors, $YBa_2Cu_3O_y$ (YBCO) and $YBa_2Cu_4O_8$ (Y124), as $\kappa$ vs $H$ up to 45 T, at two temperatures well below $T_c$. All curves exhibit the same rapid drop below a certain critical field. This is precisely the behaviour expected of a clean type-II superconductor ($l_0 >> \xi_0$), whereby the long electronic mean free path $l_0$ in the normal state is suddenly curtailed when vortices appear in the sample and scatter the electrons (see Fig. S2, and associated discussion). This effect is observed in any clean type-II superconductor, as illustrated in Fig. 1e and Fig. S2. Theoretical calculations[12] reproduce well the rapid drop of $\kappa$ at $H_{c2}$ (Fig. 1e).

To confirm our interpretation that the drop in $\kappa$ is due to vortex scattering, we have measured a single crystal of the cuprate $Tl_2Ba_2CuO_{6+\delta}$ (Tl-2201) with a much shorter mean free path, such that $l_0 \sim \xi_0$. As seen in Fig. 2a, the suppression of $\kappa$ upon entering the vortex state is much more gradual than in the ultraclean YBCO. The contrast between Tl-2201 and YBCO mimics the behavior of the type-II superconductor $KFe_2As_2$ as the sample goes from clean ($l_0 \sim 10~\xi_0$) (ref. 13) to dirty ($l_0 \sim \xi_0$) (ref. 14) (see Fig. 2b). We conclude that the onset of the sharp drop in $\kappa$ with decreasing $H$ in YBCO is a direct measurement of the critical field $H_{c2}$, where vortex scattering begins.

The direct observation of $H_{c2}$ in a cuprate material is our first main finding. We obtain $H_{c2} = 22 \pm 2$ T at $T = 1.8$ K in YBCO (at $p = 0.11$) and $H_{c2} = 44 \pm 2$ T at $T = 1.6$ K in Y124 (at $p = 0.14$) (Fig. 1a), giving $\xi_0 = 3.9$ nm and 2.7 nm, respectively. In Y124, the transport mean free path $l_0$ was estimated to be roughly 50 nm (ref. 15), so that the clean-limit condition $l_0 >> \xi_0$ is indeed satisfied. Note that the specific heat is not sensitive to vortex scattering and so should not have a marked anomaly at $H_{c2}$, as indeed found in YBCO at $p = 0.1$ (ref. 10).

We can verify that our measurement of $H_{c2}$ in YBCO is consistent with existing thermodynamic and spectroscopic data by computing the condensation energy $\delta E = H_c^2 / 2\mu_0$, where $H_c^2 = H_{c1} H_{c2} / (\ln \kappa_{GL} + 0.5)$, with $H_{c1}$ the lower critical field and $\kappa_{GL}$ the Ginzburg-Landau parameter (ratio of penetration depth to coherence length). Magnetization data[6] on YBCO give $H_{c1} = 24 \pm 2$ mT at $T_c = 56$ K. Using $\kappa_{GL} = 50$ (ref. 6), our value of $H_{c2} = 22$ T (at $T_c = 61$ K) yields $\delta E / T_c^2 = 13 \pm 3$ J / $K^2$ $m^3$. For a $d$-wave superconductor, $\delta E = N_F \Delta_0^2 / 4$, where $\Delta_0 = \alpha k_B T_c$ is the gap maximum and $N_F$ is the density of states at the Fermi energy, related to the electronic specific heat coefficient $\gamma_N = (2\pi^2/3) N_F k_B^2$, so that $\delta E / T_c^2 = (3\alpha^2 / 8\pi^2) \gamma_N$. Specific heat data[10] on YBCO at $T_c = 59$ K give $\gamma_N = 4.5 \pm 0.5$ mJ / $K^2$ mol ($43 \pm 5$ J / $K^2$ $m^3$) above $H_{c2}$. We therefore obtain $\alpha = 2.8 \pm 0.5$, in good agreement with estimates from spectroscopic measurements on a variety of hole-doped cuprates, which yield $2\Delta_0 / k_B T_c \sim 5$ between $p = 0.08$ and $p = 0.24$ (ref. 16). This shows that the value of $H_{c2}$ measured by thermal conductivity provides quantitatively coherent estimates of the condensation energy and gap magnitude in YBCO.

The position of the rapid drop in $\kappa$ vs $H$ does not shift appreciably with temperature up to $T \sim 10$ K or so (Figs. 1b and 1d), showing that $H_{c2}(T)$ is essentially flat at low temperature. This is in sharp contrast with the resistive transition at $H_{vs}(T)$, which moves down rapidly with increasing temperature (Fig. 1f). In Fig. 3, we plot $H_{c2}(T)$ and $H_{vs}(T)$ on an $H$-$T$ diagram, for both YBCO and Y124. In both cases, we see that $H_{c2} = H_{vs}$ in the $T = 0$ limit. This is our second main finding: there is no vortex liquid regime at $T = 0$. Of course, with increasing temperature the vortex-liquid phase grows rapidly, causing $H_{vs}(T)$ to fall below $H_{c2}(T)$. The same behaviour is seen in Tl-2201 (Fig. 2d): at low temperature, $H_{c2}(T)$ determined from $\kappa$ is flat whereas $H_{vs}(T)$ from resistivity falls abruptly, and $H_{c2} = H_{vs}$ at $T \rightarrow 0$ (see also Figs. S3 and S4).

Having established that $H_{c2} = H_{vs}$ at $T \to 0$ in YBCO, Y124 and Tl-2201, we can determine how $H_{c2}$ varies with doping in YBCO from measurements of $H_{vs}(T)$ (as in Figs. S5 and S6). For $p < 0.15$, fields lower than 60 T are sufficient to suppress $T_c$ to zero, and thus directly access $H_{vs}(T \to 0)$, yielding $H_{c2} = 24 \pm 2$ T at $p = 0.12$ (Fig. 3c), for example. For $p > 0.15$, however, $T_c$ cannot be suppressed to zero with our maximal available field of 68 T (Figs. 3d and S5), so an extrapolation procedure must be used to extract $H_{vs}(T \to 0)$. Following ref. 17, we obtain $H_{vs}(T \to 0)$ from a fit to the theory of vortex-lattice melting[8], as illustrated in Fig. 3 (and Fig. S6). In Fig. 4a, we plot the resulting $H_{c2}$ values as a function of doping, listed in Table S1, over a wide doping range from $p = 0.05$ to $p = 0.205$. This brings us to our third main finding: the $H$ - $p$ phase diagram of superconductivity consists of two peaks, located at $p_1 \sim 0.08$ and $p_2 \sim 0.18$. (A plot of $H_{vs}(T \to 0)$ vs $p$ was reported earlier on the basis of $c$-axis resistivity measurements[17], in excellent agreement with our own results, but the two peaks where not observed because the data were limited to $0.078 < p < 0.162$.) The two-peak structure is also apparent in the usual $T$ - $p$ plane: the single $T_c$ dome at $H = 0$ transforms into two domes when a magnetic field is applied (Fig. 4b).

A natural explanation for two peaks in the $H_{c2}$ vs $p$ curve is that each peak is associated with a distinct critical point where some phase transition occurs[18]. An example of this is the heavy-fermion metal CeCu$_2$Si$_2$, where two $T_c$ domes in the temperature-pressure phase diagram were revealed by adding impurities to weaken superconductivity[19]: one dome straddles an underlying antiferromagnetic transition and the other dome a valence transition. In YBCO, there is indeed strong evidence of two transitions – one at $p_1$ and another at a critical doping consistent with $p_2$ (ref. 20). In particular, the Fermi surface of YBCO is known to undergo one transformation at $p = 0.08$ and another near $p \sim 0.18$ (ref. 4). Hints of two critical points have also been found in Bi$_2$Sr$_2$CaCu$_2$O$_{8+\delta}$, as changes in the superconducting gap detected by ARPES at $p_1 \sim 0.08$ and $p_2 \sim 0.19$ (ref. 21).

The transformation at $p_2$ is a reconstruction of the large hole-like cylinder at high doping that produces a small electron pocket[4,5,22]. We associate the fall of $T_c$ and the collapse of $H_{c2}$ below $p_2$ to that Fermi-surface reconstruction. Recent studies indicate that charge-density wave order is most likely the cause of the reconstruction[23,24,25]. Indeed, the charge modulation seen with X-rays[24,25] and the Fermi-surface reconstruction seen in the Hall coefficient[4] emerge in parallel with decreasing temperature (see Fig. S7). Moreover, the charge modulation amplitude drops suddenly below $T_c$, showing that superconductivity and charge order compete[24,25] (Fig. S8a). As a function of field[25], the onset of this competition defines a line in the $H$ - $T$ plane (Fig. S8B) that is consistent with our $H_{c2}(T)$ line (Fig. 3). The flip side of this phase competition is that superconductivity must in turn be suppressed by charge order, consistent with our interpretation of the $T_c$ fall and $H_{c2}$ collapse below $p_2$.

We can quantify the impact of phase competition by computing the condensation energy $\delta E$ at $p = p_2$, using $H_{c1} = 110 \pm 5$ mT at $T_c = 93$ K (ref. 7) and $H_{c2} = 150 \pm 20$ T, and comparing with $\delta E$ at $p = 0.11$ (see above): $\delta E$ decreases by a factor 20, and $\delta E / T_c^2$ by a factor 8. In Fig. 4c, we plot the doping dependence of $\delta E / T_c^2$ and find good qualitative agreement with earlier estimates based on specific heat data[26] (see Fig. S9 and associated discussion). The tremendous weakening of superconductivity below $p_2$ is attributable to a drop in the density of states as the large hole-like Fermi surface reconstructs into small pockets. This process may well involve both the pseudogap formation and the charge ordering.

Upon crossing below $p_1$, the Fermi surface of YBCO undergoes a second transformation, signalled by pronounced changes in transport properties[4,5] and in the effective mass $m^\star$ (ref. 27), where the small electron pocket disappears. This is strong evidence that the peak in $H_{c2}$ at $p_1 \sim 0.08$ (Fig. 4a) coincides with an underlying critical point. This critical point is presumably associated with the onset of incommensurate spin modulations detected below $p \sim 0.08$ by neutron scattering[28] and muon spectroscopy[29]. Note that the increase in $m^\star$ naturally explains the increase in $H_{c2}$ going from $p = 0.11$ (local minimum) to $p = 0.08$, since $H_{c2} \sim 1 / \xi_0{}^2 \sim 1 / v_F{}^2 \sim m^{\star 2}$.

Our findings shed light on the $H$-$T$-$p$ phase diagram of YBCO, in three different ways. In the $H$-$p$ plane, they establish the boundary of the superconducting phase and reveal a two-peak structure, the likely fingerprint of two underlying critical points. In the $H$-$T$ plane, they delineate the separate boundaries of vortex solid and vortex liquid phases, showing that the latter phase vanishes as $T \rightarrow 0$. In the $T$-$p$ plane, they elucidate the origin of the dome-like $T_c$ curve as being due primarily to phase competition, rather than phase fluctuations, and quantify the impact of that competition on the condensation energy.

---

## Author contributions

G.G., S.R.d.C. and N.D.-L. performed the thermal conductivity measurements at Sherbrooke. G.G., O.C.-C., S.D.-B., S.K. and N.D.-L. performed the thermal conductivity measurements at the LNCMI in Grenoble. G.G., O.C.-C., A.J.-F., D.G. and N.D.-L. performed the thermal conductivity measurements at the NHMFL in Tallahassee. N.D.-L., D.L., M.S., B.V. and C.P. performed the resistivity measurements at the LNCMI in Toulouse. S.R.d.C., J.C., J.-H.P. and N.D.-L. performed the resistivity measurements at the NHMFL in Tallahassee. M.-È.D., O.C.-C., G.G., F.L., D.L. and N.D.-L. performed the resistivity measurements at Sherbrooke. B.J.R., R.L., D.A.B. and W.N.H. prepared the YBCO and Tl-2201 single crystals at UBC (crystal growth, annealing, de-twinning, contacts). S.A. and N.E.H. prepared the Y124 single crystals. G.G., O.C.-C., F.L., N.D.-L. and L.T. wrote the manuscript. L.T. supervised the project.

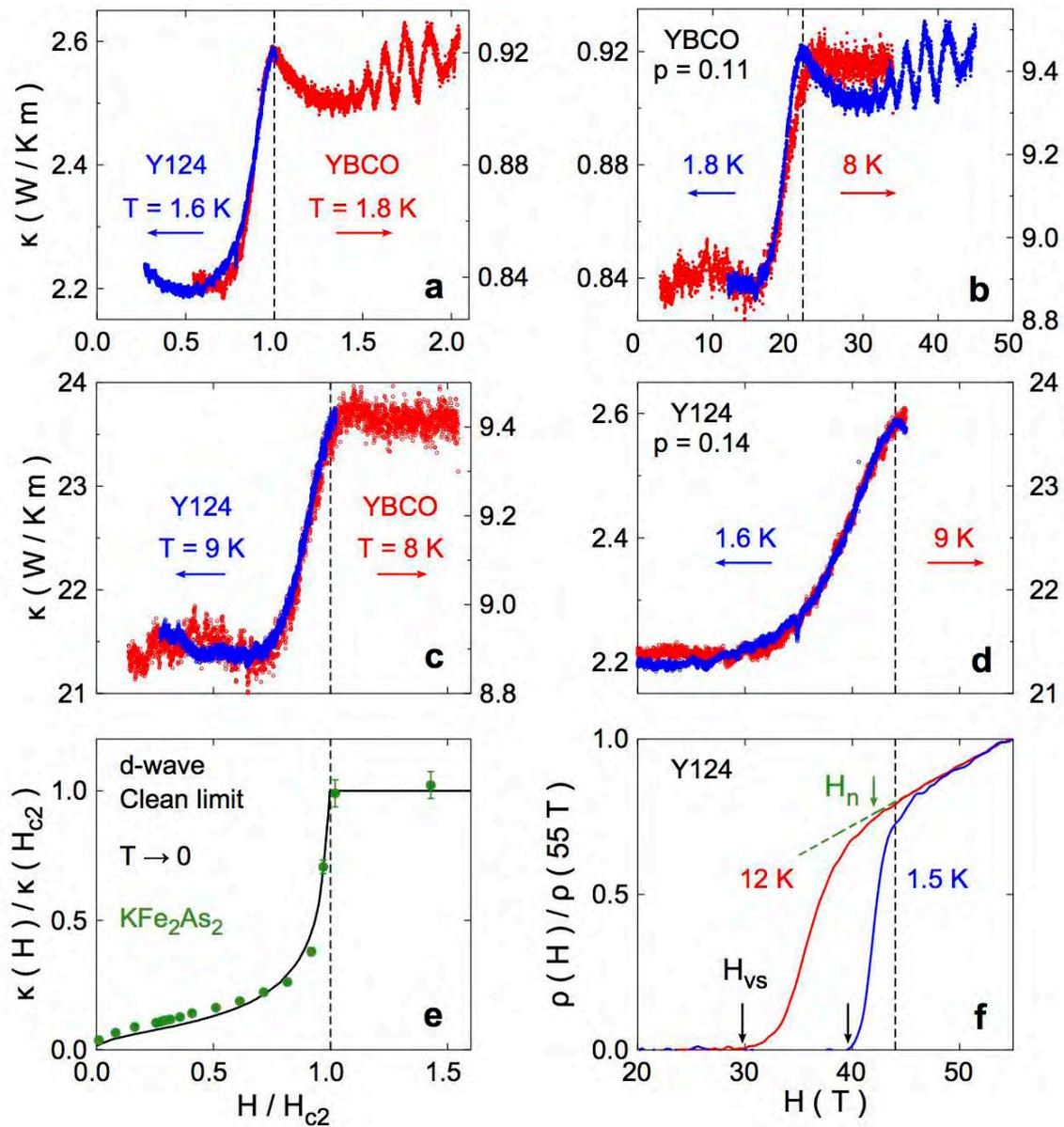

**Figure 1 | Field dependence of thermal conductivity.**

**a), b), c), d)** Magnetic field dependence of the thermal conductivity κ in YBCO ($p$ = 0.11) and Y124 ($p$ = 0.14), for temperatures as indicated. The end of the rapid rise marks the end of the vortex state, defining the upper critical field $H_{c2}$ (vertical dashed line). In Figs. 1a and 1c, the data are plotted as κ vs $H / H_{c2}$, with $H_{c2}$ = 22 T for YBCO and $H_{c2}$ = 44 T for Y124. The remarkable similarity of the normalized curves demonstrates the good reproducibility across dopings. The large quantum oscillations seen in the YBCO data above $H_{c2}$ confirm the long electronic mean path in this sample.  In Figs. 1b and 1d, the overlap of the two isotherms plotted as $\kappa$ vs $H$ shows that $H_{c2}(T)$ is independent of temperature in both YBCO and Y124, up to at least 8 K. **e)** Thermal conductivity of the type-II superconductor $KFe_2As_2$ in the $T = 0$ limit, for a sample in the clean limit (green circles). The data[13] are compared to a theoretical calculation for a $d$-wave superconductor in the clean limit[12]. **f)** Electrical resistivity of Y124 at $T$ = 1.5 K (blue) and $T$ = 12 K (red) (ref. 15). The green arrow defines the field $H_n$ below which the resistivity deviates from its normal-state behaviour (green dashed line). While $H_{c2}(T)$ is essentially constant up to 10 K (Fig. 1d), $H_{vs}(T)$ – the onset of the vortex-solid phase of zero resistance (black arrows) – moves down rapidly with temperature (see also Fig. 3b).

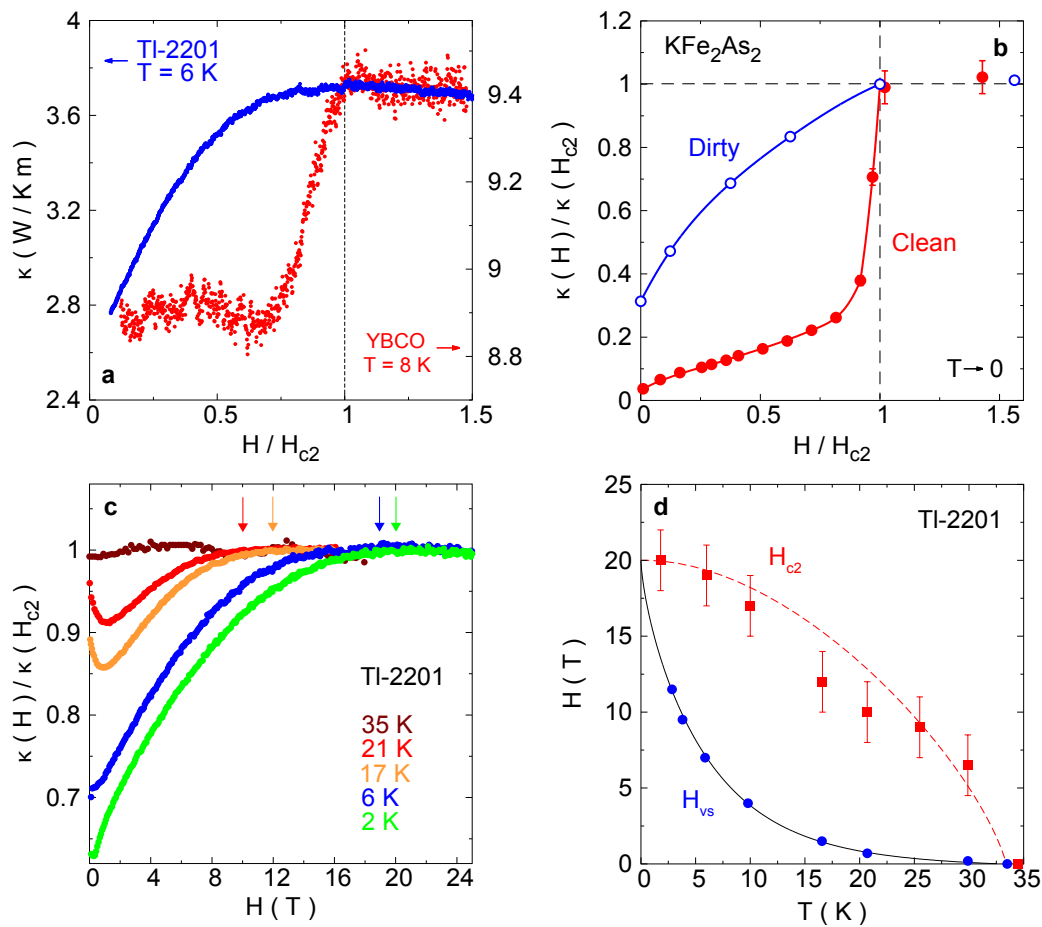

**Figure 2 | Thermal conductivity of TI-2201.**

**a)** Magnetic field dependence of the thermal conductivity κ in TI-2201, measured at $T$ = 6 K on an overdoped sample with $T_c$ = 33 K (blue). The data are plotted as κ vs $H / H_{c2}$, with $H_{c2}$ = 19 T, and compared with data on YBCO at $T$ = 8 K (red; from Fig. 1b), with $H_{c2}$ = 23 T. **b)** Corresponding data for KFe$_2$As$_2$, taken on clean[13] (red) and dirty[14] (blue) samples. **c)** Isotherms of κ( $H$ ) in TI-2201, at temperatures as indicated, where κ is normalized to unity at $H_{c2}$ (arrows). $H_{c2}$ is defined as the field below which κ starts to fall with decreasing field. **d)** Temperature dependence of $H_{c2}$ (red squares) and $H_{vs}$ (blue circles) in TI-2201. The error bars reflect the uncertainty in locating the drop in κ vs $H$. All lines are a guide to the eye.

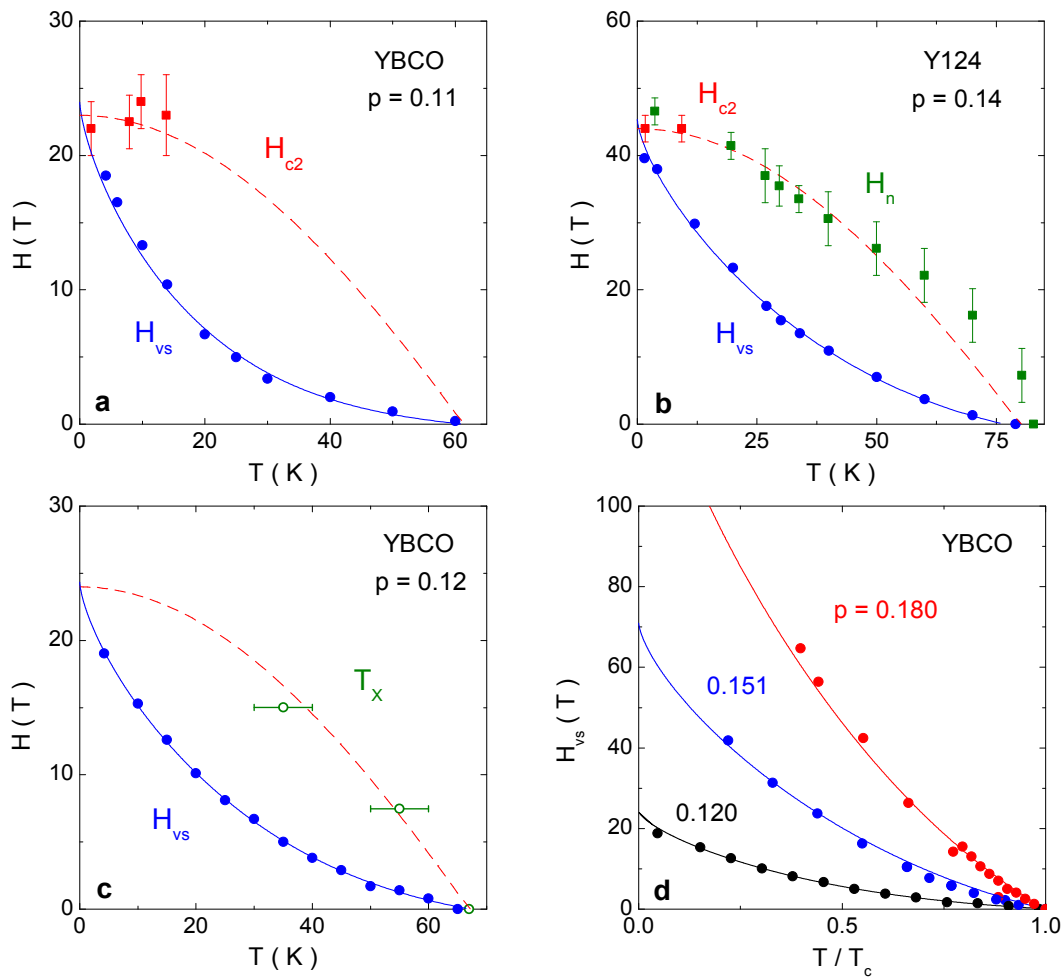

**Figure 3 | Field-temperature phase diagrams.**

**a), b)** Temperature dependence of $H_{c2}$ (red squares, from data as in Fig. 1) for YBCO and Y124, respectively. The red dashed line is a guide to the eye, showing how $H_{c2}(T)$ might extrapolate to zero at $T_c$. The solid lines are a fit of the $H_{vs}(T)$ data (solid circles) to the theory of vortex-lattice melting[8], as in ref. 17. Note that $H_{c2}(T)$ and $H_{vs}(T)$ converge at $T = 0$, in both materials, so that measurements of $H_{vs}$ vs $T$ can be used to determine $H_{c2}(0)$ in YBCO. In Fig. 3b, we plot the field $H_n$ defined in Fig. 1f (open green squares, from data in ref. 15), which corresponds roughly to the upper boundary of the vortex-liquid phase (see Supplementary Material). We see that $H_n(T)$ is consistent with $H_{c2}(T)$.

**c)** Temperature $T_X$ below which charge order is suppressed by the onset of superconductivity in YBCO at $p$ = 0.12, as detected by X-ray diffraction[25] (open green circles, from Fig. S8). We see that $T_X(H)$ follows a curve (red dashed line) that is consistent with $H_n(T)$ (at $p$ = 0.14; Fig. 1f) and with the $H_{c2}(T)$ detected by thermal conductivity at lower temperature (at $p$ = 0.11 and 0.14). **d)** $H_{vs}(T)$ vs $T / T_c$, showing a dramatic increase in $H_{vs}(0)$ as $p$ goes from 0.12 to 0.18. From these and other data (in Fig. S6), we obtain the $H_{vs}(T{\rightarrow}0)$ values that produce the $H_{c2}$ vs $p$ curve plotted in Fig. 4a.

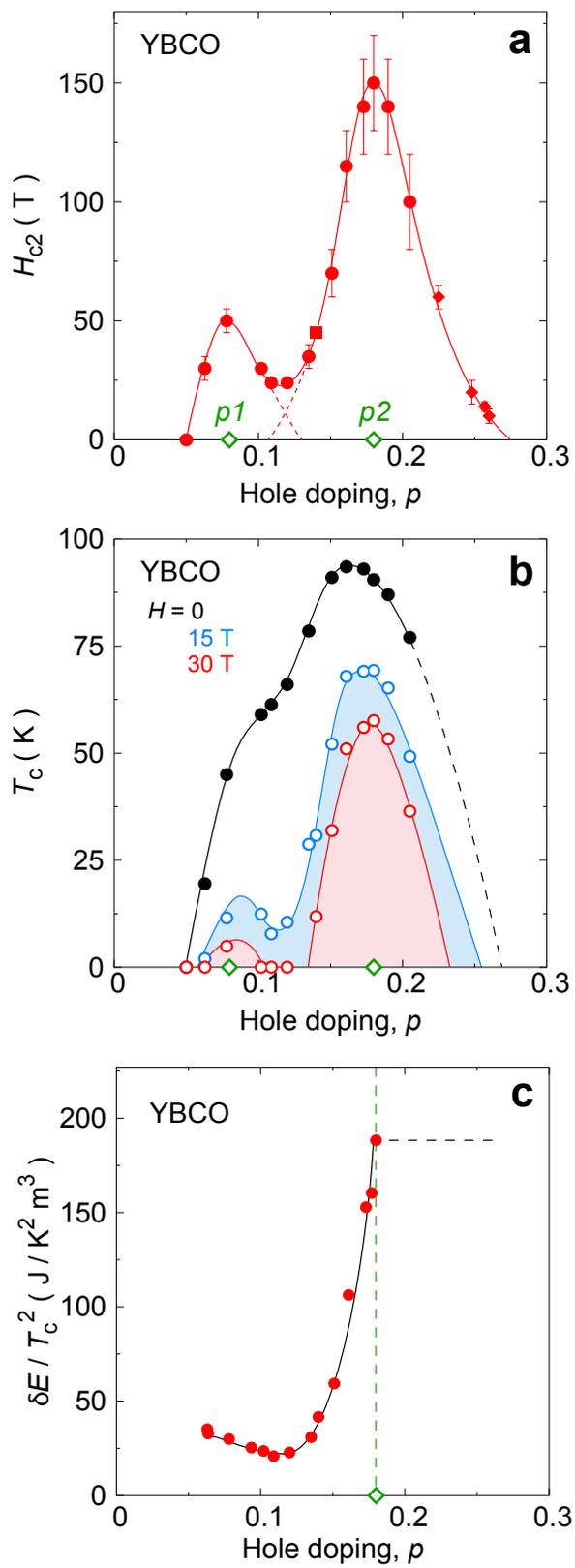

**Figure 4 | Doping dependence of $H_{c2}$, $T_c$ and the condensation energy.**

**a)** Upper critical field $H_{c2}$ of the cuprate superconductor YBCO as a function of hole concentration (doping) $p$. $H_{c2}$ is defined as $H_{vs}(T{\to}0)$ (Table S1), the onset of the vortex-solid phase at $T \to 0$, where $H_{vs}(T)$ is obtained from high-field resistivity data (Figs. 3, S5 and S6). The point at $p = 0.14$ (square) is from data on Y124 (Fig. 3b). **b)** Critical temperature $T_c$ of YBCO as a function of doping $p$, for three values of the magnetic field $H$, as indicated (Table S1). $T_c$ is defined as the point of zero resistance. All lines are a guide to the eye. Two peaks are observed in $H_{c2}(p)$ and in $T_c(p; H > 0)$, located at $p_1 \sim 0.08$ and $p_2 \sim 0.18$ (open diamonds). The first peak coincides with the onset of incommensurate spin modulations at $p \approx 0.08$, detected by neutron scattering[28] and muon spin spectroscopy[29]. The second peak coincides with the approximate onset of Fermi-surface reconstruction[4,5], attributed to charge modulations detected by high-field NMR (ref. 23) and X-ray scattering[24,25]. **c)** Condensation energy $\delta E$ (full red circles), given by the product of $H_{c2}$ and $H_{c1}$ (see text and Fig. S9), plotted as $\delta E \, / \, T_c^2$. Note the 8-fold drop below $p_2$ (vertical dashed line), attributed predominantly to a corresponding drop in the density of states.